\documentstyle[12pt,aaspp4]{article}

\def\ep{\epsilon}

\def\ni{\noindent}

\def\bar{\overline}

\def\OB{\overline{\bf B}}
\def\emf{\overline{\mbox{${\cal E}$}} {}}

\def\emfb{\overline{\mbox{\boldmath ${\cal E}$}} {}}
\def\emfb{\overline{\mbox{\boldmath ${\cal E}$}} {}}

\def\bbE{\overline {\bf E}}
\def\bbh{\overline {\bf h}}

\def\beq{\begin{equation}}
\def\ee{\end{equation}}
\def\lsim{\mathrel{\rlap{\lower4pt\hbox{\hskip1pt$\sim$}}
    \raise1pt\hbox{$<$}}}
\def\gsim{\mathrel{\rlap{\lower4pt\hbox{\hskip1pt$\sim$}}
    \raise1pt\hbox{$>$}}}
\def\bfE{{\bf E}}
\def\bfJ{{\bf J}}

\def\bfA{{\bf A}}
\def\bfa{{\bf a}}
\def\bfe{{\bf e}}
\def\bfB{{\bf B}}
\def\bbJ{\bar {\bf J}}

\def\bB{\overline B}

\def\bV{\overline V}
\def\ts{\times}
\def\lb{\langle}
\def\rb{\rangle}
\def\curl{\nabla {\ts}}

\def\bbV{\bar {\bf V}}

\def\bfv{{\bf v}}

\def\bfV{{\bf V}}
\def\bfj{{\bf j}}

\def\bfe{{\bf e}}

\def\bfb{{\bf b}}

\def\bfB{{\bf B}}

\def\bbB{\overline {\bf B}}
\def\bbA{\overline {\bf A}}

\def\div{\nabla\cdot}

\begin{document}
\title{Laboratory Plasma  Dynamos, Astrophysical Dynamos,
and Magnetic Helicity Evolution}
\author{Eric G. Blackman\altaffilmark{1,2}, Hantao Ji\altaffilmark{3}}
\affil{1. Dept. of Physics and Astronomy, Univ. of Rochester,
    Rochester, NY, 14627, USA; 2. Laboratory for Laser Energetics Univ. of Rochester, Rochester NY, 14623, USA; 3. Center for Magnetic Self-organization of Laboratory and Astrophysical Plasmas, Princeton Plasma Physics Laboratory, P.O. Box 451, Princeton NJ 08543}


\begin{abstract}
The  term ``dynamo'' means different things to 
 the laboratory fusion plasma  and astrophysical plasma communities.
To alleviate the resulting confusion and to  
facilitate interdisciplinary progress, 
we pinpoint  conceptual differences and similarities
between  laboratory plasma dynamos 
and astrophysical dynamos.  We can divide dynamos into three types: 1.
magnetically dominated helical dynamos
which sustain a large scale magnetic field against resistive decay
and drive the magnetic geometry toward the lowest energy state,
2. flow-driven helical dynamos which amplify or sustain large
scale magnetic fields in an otherwise turbulent flow, and 
3. flow-driven nonhelical dynamos which amplify 
fields on scales at or below the driving turbulence.
We discuss how all three types occur in astrophysics whereas 
plasma confinement device dynamos are of the first type.
Type 3 dynamos requires no magnetic or kinetic helicity of any kind.
Focusing on type 1 and 2 dynamos, 
we show how different limits of a unified set of 
equations for magnetic helicity evolution 
reveal both types. 
We explicitly describe  a steady-state  example of a type 1 dynamo,
and three examples of type 2 dynamos: (i) closed volume and time dependent; 
(ii) steady-state with open boundaries; (iii) time dependent with open boundaries.

\end{abstract}


\ni {\bf Key Words:} magnetic fields;  MHD; 
Sun: coronal mass ejections;  stars: coronae;
accretion, accretion disks;methods: laboratory;

\section{Introduction}

\subsection{General Motivation}
Dynamos describe the amplification and/or 
 sustenance  of magnetic fields in electrically
conducting media such as plasmas and liquid metals.
For 50 years the detailed dynamics and mechanisms of dynamos have 
been subjects of active research  in  plasma physics, astrophysics,
geophysics, and nonlinear dynamics. Historically, 
much dynamo research in these fields has  progressed independently.
Unsurprisingly, when different communities interact, 
confusion arises  when the term ``dynamo'' is used. This is partly 
because  laboratory plasmas are typically magnetically dominated  
whereas the interiors of astrophysical rotators are typically flow dominated.
 Astrophysicists
familiar only with dynamos in flow dominated environments 
wonder what role a dynamo could possibly play in a magnetically dominated
environment as they are used to thinking of 
a dynamo as a flow driven amplification of an initially weak magnetic
field.

Presently, there is timely motivation to alleviate this confusion 
and  increased opportunity for interdisciplinary research.  
Flow driven dynamos and MHD instabilities are being  
increasingly studied in liquid metals (\cite{gailitisa,gailitisb,jimri,noguchi,peffley,sisan}
to address some principles
of traditional astrophysical and planetary
dynamos,  and the magnetically dominated
dynamos of confinement plasmas 
are now realized to have direct analogies 
in astrophysical coronae (e.g. \cite{ji04,blackman05}).

Generally, it is important to clarify what is meant by a dynamo in each context
so that all research communities can appreciate the common
principles and differences.
A particular unifying question for dynamo theories  is 
to what extent are helical dynamos 
independent of the  resistivity (or dissipation).
While different answers  to this question arise in 
different contexts, we aim to guide the reader toward understanding
these different answers in a unified framework that draws from recent
work on magnetic helicity evolution.

\subsection{Distinguishing Dynamos}

Laboratory plasma dynamos arise in magnetically dominated conditions 
that have been studied in the context of  fusion plasma confinement,  
such as the Reversed Field Pinch (RFP)
configuration (\cite{bodin90,jiprager}). These dynamos 
describe how a large scale 
magnetic configuration adjusts toward its relaxed state, 
in the presence of external driving away from the relaxed state
by a magnetic field aligned electric field
(or equivalently, external injection of one sign of magnetic helicity) 
(\cite{ji99,jiprager,ortolani93,strauss85,strauss86,bhattacharjee86,holmesetal88,gd1,by,bellan}).
The magnetic helicity injection actually 
drives the system away from the relaxed state,
but also generates small amplitude fluctuation via 
kink mode instabilities from large currents, or 
tearing modes from large current gradients. The fluctuations produce a 
correlation between 
fluctuating velocity and magnetic fields--- 
the turbulent electromotive force (EMF) 
$\emfb=\lb\bfv\ts\bfb\rb$, which
allows the system to evolve back toward a relaxed state.
The turbulent EMF reduces the field aligned current
to restore stability by driving a spatial flow of magnetic helicity.
This enables the magnetic field structure to
evolve toward the largest helical scale available (subject to boundary
conditions), as this is the lowest energy state (\cite{taylor86}).
Continuous injection of magnetic helicity 
typically leads to a quasi-steady dynamical equilibrium 
 with (sawtooth type) oscillations
as the system is driven away from, and then evolves back toward the relaxed
state. 
If the injection is turned off, the fully  
relaxed state can be reached, but the field
eventually resistively decays.
Via the dynamo, the injection  therefore also sustains  
the large scale helical field against  decay.

In short, laboratory plasma dynamos involve: 
(1) external injection of one sign of magnetic helicity
(2) a change the magnetic field structure 
with conversion of magnetic flux from
toroidal to poloidal (or vice versa) geometries,
(3) an increase the scale of the field as the scale of magnetic
helicity increases in the relaxation phase,
and (4) sustenance of helical field against
dissipation.
Boundary value studies have focused on all of the above, as well as 
the specific instabilities that drive the small amplitude fluctuations.
Laboratory plasma dynamos involve both favorable and unfavorable features for
plasma confinement: On the one hand they sustain a large scale field
in an ordered configuration. But
to sustain this state,  instabilities are required, which 
produce unwanted dissipation and heat transport.

The magnetically dominated dynamos of laboratory plasmas can be contrasted to the mean field helical dynamo originally applied
inside of flow dominated astrophysical rotators 
(\cite{moffatt,parker,krause,zeldovich83}).
The latter involve an initially
weak large scale field which is subsequently amplified via strong, large amplitude helical velocity fluctuations. (In this context, fluctuations and
turbulence are used interchangeably.).
That laboratory plasma dynamos involve small amplitude fluctuations
around a strong large scale field implies that high order
correlations of fluctuations can be justifiably ignored.
In contrast, the fact that  fluctuations typically
dominate in a velocity driven dynamo 
requires  a more sophisticated closure for a rigorous theory
because high order correlations of fluctuations cannot be 
straightforwardly  neglected.

An important feature of velocity driven helical dynamos is that the
large scale field is amplified and sustained
on scales significantly larger than the scale of the driving turbulence.
This contrasts the magnetically driven 
laboratory plasma dynamo, where the fluctuations 
vary on time scales short compared to the 
mean field evolution but can be large scale
in space. The key similarity between the laboratory plasma dynamos and the 
flow-driven helical dynamos is that both thrive from a 
finite $\emfb\cdot \bbB$. This quantity drives the
large scale helical field relaxation in the laboratory plasma dynamo and  
amplifies large scale helical fields in the flow-driven case.
For the latter, the source of $\emfb_{||}$ (where $||$ indicates along $\bbB$)
typically, though not  exclusively, is the 
kinetic helicity, $\lb\bfv\cdot\curl\bfv\rb$, 
a pseudoscalar correlation arising from the interplay
between stratified turbulence and rotation. For a real system with an
outward decreasing density, this can be sustained as rising eddies expand 
and rotate oppositely to the underlying mean rotation to conserve angular
momentum. Falling eddies rotate in the same direction as the 
mean rotation.  Both rising and falling eddies thus
statistically contribute the same sign of kinetic helicity (or twist)
$\lb\bfv \cdot \curl \bfv\rb$.
The northern and southern hemispheres
have opposite signs of this helicity.
Inside astrophysical rotators, the ultimate source of energy for the turbulence
is typically convection or differential rotation. The magnetic
field is not dominant therein but the turbulence
which facilitates the large scale field growth is necessarily accompanied
by magnetic fluctuations with energy density
comparable to that in the velocity fluctuations.
Recently, boundary terms, helicity fluxes and anisotropic
contributions  from mean velocity flows 
 have been considered
as additional contributions to the electromotive force driving
the velocity driven helical dynamo.
(e.g. Blackman \& Field 2000a; Vishniac \& Cho 2001; R\"adler et al. 2003,
Subramanian \& Brandenburg 2004; 
Brandenburg \& Subramanian 2005).  

The growth of small scale magnetic fluctuations in the velocity driven
dynamo environments of astrophysical rotators 
also highlights another important concept that distinguishes flow
driven dynamos from the magnetically driven dynamos in laboratory plasmas:
laboratory dynamos always involve helicity, whereas both
helical and non-helical 
 flow-driven dynamos exist. 
Non-helical dynamos (\cite{kazanstev},\cite{maroncowley},
\cite{haugen03},\cite{haugen04},
\cite{schek02},\cite{schek04}) 
do not involve a mean turbulent electromotive force,
just a turbulent velocity  which amplifies
magnetic energy via random walk line stretching and shear.
Non-helical and helical flow-driven  dynamos also differ in that
the former amplifies magnetic energy only up to
the input driving scale, whereas the latter
can amplify fields on even larger scales, as needed
to explain observed large scale dynamo cycle periods
in astrophysical objects.
An important complication is that small scale dynamos 
can operate concurrently with the large scale helical dynamo, 
and the effect of the small scale field growth on the large scale dynamo
has been a subject of considerable research.

While astrophysical dynamos are typically thought of as flow dominated, 
it is important to note that 
coronae above astrophysical rotators  such as stars and accretion disks,
are likely magnetically dominated (e.g.\cite{galeev79},\cite{sz},\cite{hm93},
\cite{fieldrogers93}).
Therefore, astrophysical coronae are in fact sites for 
magnetically dominated dynamos driven by helical
field injection from the astrophysical rotator below (\cite{bf04,blackman05}).
For all extra-terrestrial astrophysical rotators except galaxies,
we observe at most the coronal fields, not the interior field.
In the sun, stars, and accretion engines there is  evidence
for the presence of large scale coronal magnetic fields.
For example, the coronal holes of the sun are sites of large scale ``open'' 
field lines along
which the solar wind propagates (e.g. \cite{sz}). 
The reversal in sign of these fields
indicates that they need to be regenerated every 11 years.
Analogously, the jets from accretion engines such as 
young stellar objects, active galactic nuclei and even 
gamma-ray bursts are likely 
magnetically mediated either from fling type models
(see e.g. \cite{pudritz04})
or magnetic towers (e.g. \cite{lb03,um06}).
The needed large scale coronal fields
can be produced by the opening up of smaller scale
loops from within the rotator below, as seen in the sun (Wang \& Sheeley 2003).
These circumstances reveal that magnetically
driven helical dynamos, in addition to the flow-driven helical dynamo also 
play an important  role in astrophysical magnetic field evolution.
Indeed, the most direct analogy of laboratory plasma
dynamos in astrophysics is the magnetic field relaxation in 
astrophysical coronae subject to helicity injection at the base.
The injection drives the system away from the relaxed state but instabilities
arise, allowing the system to relax. Steady injection wold lead
to a quasi-steady dynamical equilibrium.

Recent progress toward understanding helical dynamos 
has resulted from a combination of numerical
and analytic work that dynamically 
incorporates magnetic helicity evolution 
(\cite{pfl}, \cite{kleeorin82}, Ji 1999, Field \& Blackman 2002
\cite{kr}, \cite{bf00a},  
\cite{rk}, \cite{b2001}, \cite{bf02}, Vishniac \& Cho 2002, \cite{maronblackman}, 
\cite{bb02}, \cite{kleeorin02}, \cite{bb03}, \cite{blackman03}, \cite{sb04}, 
 \cite{bs05}).  However, identification of common principles 
for laboratory plasma dynamos, helical flow-driven dynamos, and magnetically
dominated coronal relaxation  
has been obscured by the particular approximations made or by the details
of the application.
The purpose of this paper is to show how a range of different dynamo
theories  incorporate magnetic helicity conservation 
and how they can be seen to represent different cases of
a unified framework.  We derive
the simplest version of the equations representative of a particular
type of helical dynamo. We purposely do not present complete
analyses of the solutions  (see \cite{bs05} for a review); 
the main goal here is to identify where
the equations  fit into the big picture.

In section 2 we derive the general magnetic helicity density
evolution equations for a two-scale system including time
dependent and boundary terms.  In section 3 we discuss how  specific
cases of these equations correspond to 4 specific dynamo circumstances:
(1) a steady-state magnetically driven laboratory plasma dynamo, (2)
a time-dependent closed flow-driven dynamo,
(3) a steady state open flow driven dynamo and its implications
for subsequent magnetically driven coronal magnetic relaxation (4) an open
time dependent flow-driven dynamo.  
In section 4 we discuss insights about  quenching and steady states
that emerge from the unified framework.
We  conclude in section 5.

\section{Unifying Equations for Helical Dynamos}

For completeness we derive the needed magnetic helicity
evolution equations and consolidate them into a useful form
for all subsequent sections of this paper.
(See also e.g. \cite{bellan}).

We start with the electric field 
\beq
\bfE=-\nabla\Phi -\partial_t\bfA,
\label{1}
\ee
where $\Phi$ and $\bfA$ are the scalar and vector potentials.
Taking the average (spatial, temporal, or ensemble) 
and denoting averaged quantities by an overbar 
we obtain 
\beq
\bbE=-\nabla{\overline \Phi} -\partial_t\bbA
\label{2}
\ee
Subtracting (\ref{2}) from (\ref{1}) gives the equation
for the fluctuating electric field
\beq
\bfe=-\nabla\phi -\partial_t\bfa
\label{3},
\ee
where $\phi$ and $\bfa$ are the fluctuating scalar and vector potentials.

Now, using 
$\bfB\cdot \partial_t \bfA= \partial_t(\bfA\cdot \bfB) +\bfE\cdot \bfB -\nabla \cdot (\bfA\ts \bfE)$,
where the latter two terms result 
from using Maxwell's equation $\partial_t \bfB=-\curl \bfE$,
and the vector identity 
$\bfA \cdot \curl \bfE = \bfE\cdot\bfB-\nabla \cdot (\bfA \ts \bfE)$, 
we take the dot product of (\ref{1}) with $\bfB$ to obtain
\beq
\partial_t(\bfA\cdot\bfB)= -2(\bfE\cdot\bfB)
-\div(\Phi \bfB + \bfE\ts \bfA)
=-2(\bfE\cdot\bfB)
-\div( 2\Phi \bfB + \bfA \ts \partial_t\bfA).
\label{4}
\ee
Eq. (\ref{4}) describes the time evolution of magnetic helicity density.
The same procedure used to derive (\ref{4}) can be used, after 
dotting (\ref{2}) and (\ref{3}) 
with $\bbB$ and $\bfb$ respectively, to obtain 
equations for the time evolution of the mean magnetic helicity density
\beq
\partial_t(\bbA\cdot\bbB)= -2\bbE\cdot\bbB
-\div ({\overline\Phi} \bbB + \bbE\ts \bbA)
=-2\bbE\cdot\bbB
-\div( 2{\overline \Phi} \bbB + \bbA \ts \partial_t\bbA),
\label{5}
\ee
and fluctuating magnetic helicity density
\beq
\partial_t\overline{\bfa\cdot\bfb}= -2\overline{\bfe\cdot\bfb}
-\div(\overline{{\phi} \bfb} + \overline{\bfe\ts \bfa})
=-2\overline{\bfe\cdot\bfb}
-\div (2\overline{ \phi \bfb} + \overline{\bfa\ts \partial_t\bfa}).
\label{6}
\ee

We can eliminate the electric fields 
from (\ref{4}-\ref{6}) 
by using Ohm's law.  Here we consider  the basic Ohm's law
with only a resistive term. For the total electric field, we have
\beq
{\bfE}=-\bfV\ts\bfB +\eta \bfJ.
\label{7}
\ee
Taking the average gives the mean field Ohm's law
\beq
{\bbE}=-\emfb -\bbV\ts\bbB+\eta \bbJ,
\label{8}
\ee
where $\emfb\equiv \overline{\bfv\ts\bfb}$ is the turbulent electromotive
force.
Subtracting (\ref{8}) from (\ref{7}) gives  Ohm's law for the
fluctuating field
\beq
{\bfe}=\emfb-\bfv\ts\bfb -\bfv\ts\bbB -\bbV\ts\bfb 
+\eta \bfj.
\label{9}
\ee
Plugging (\ref{7}) into (\ref{4}), gives 
\beq
\partial_t( \bfA\cdot\bfB)= -2\eta (\bfJ\cdot\bfB)
-\div(\Phi \bfB + \bfE\ts \bfA)
=-2\eta (\bfJ\cdot\bfB)
-\div( 2\Phi \bfB + \bfA \ts \partial_t\bfA).
\label{4a}
\ee
Plugging  (\ref{8}) into (\ref{5}) and (\ref{9}) into (\ref{6}) give,
respectively, 
\beq
\partial_t(\bbA\cdot\bbB)= 
2\emfb\cdot\bbB
-2\eta \bbJ\cdot\bbB
-\div(\Phi \bbB + \bbE\ts \bbA)
=2\emfb\cdot\bbB
-2\eta \bbJ\cdot\bbB
-\div( 2{\overline \Phi} \bbB + \bbA \ts \partial_t\bbA)
\label{5a}
\ee
and fluctuating magnetic helicity
\beq
\partial_t\overline{ \bfa\cdot\bfb}= 
-2\emfb\cdot\bbB-
2\eta\overline{\bfj\cdot\bfb}
-\div(\overline{\phi \bfb} + \overline{\bfe\ts \bfa})
=-2\emfb\cdot\bbB-
2\eta\overline{\bfj\cdot\bfb}
-\div(\overline{2{\phi} \bfb} + \overline{\bfa\ts \partial_t\bfa}).
\label{6a}
\ee
Dotting (\ref{9}) with $\bfb$ and averaging 
reveals the important relation
\beq
{\emfb\cdot\bbB}=\overline{\bfe\cdot\bfb}-\eta\overline{\bfj\cdot\bfb}.
\label{13}
\ee
All helical dynamos  require  a finite $\emfb\cdot\bbB$.
Eqs. (\ref{5a}-\ref{13}) 
are the key equations to be
used in subsequent sections.

\section{Examples of Dynamos from Magnetic Helicity Evolution}

Here we show how different types of dynamos can be understood 
as limiting cases of the equations in the previous 
section.

\subsection{Steady-State, Magnetically Dominated Laboratory Plasma Dynamo}

In this subsection we work in the context of a torus. We  
take mean quantities to be time averages
and  spatial averages over  periodic directions $\phi$ (locally $\hat{\bf z})$ and $\theta$,
but not over  radius $r$ (where $r=0$ is corresponds to an azimuthal ring 
at the center of the torus' cross section.) 
While realistic RFPs and Tokamaks are time dependent,
to illustrate the relevant dynamo  simply,  
we assume mean quantities are  
steady, and that fluctuations vary on time scales much less
than the averaging time. For RFPs,
sawtooth oscillations and crashes occur over millisecond time scales 
and fluctuations occur in $100\mu s$ or shorter (e.g. \cite{jiprager},\cite{ortolani93}).
Quantities averaged over $\gsim 10$ms  can be approximated
as steady. We write the steady-state limit of (\ref{6a}) as
\beq
\emfb_{||}=
{\bbB\over \bB^2}(\nabla\cdot\bbh-\eta\overline{\bfj\cdot\bfb}),
\label{14b}
\ee
where $\bbh\equiv - \overline{
\phi\bfb}+{1\over 2}\overline{\bfa \ts\partial_t\bfa}$.
Dotting (\ref{14b}) with $\bbB$, and using (\ref{8})
gives
\beq
\emfb\cdot \bbB = 
\div {\bbh}-\eta\overline{\bfj\cdot\bfb}  
=\eta \bbJ\cdot\bbB - \bbE\cdot\bbB.
\label{new8}, 
\ee
where, from Eqn. (\ref{5}), we also have 
\beq
\bbE\cdot\bbB=
-\div( {\overline \Phi} \bbB + {1\over2}\bbA \ts \partial_t\bbA).
\label{new8bb}
\ee

The dynamo effect in  laboratory plasma configurations such as an RFP  
emerges when a large scale electric field
$\bbE$ is externally applied along the initial toroidal magnetic field 
(this represents helicity  injection usually 
via the divergence term, as emphasized  below (\ref{voltage})).
Were there no induced electromotive force $\emfb$, 
the measured current term on the right hand side of 
(\ref{new8}) would have to 
balance the applied large scale electric field along the 
magnetic field $\bbE_{||}$ in a steady state.
But for sufficiently large applied $\bbE_{||}$, 
RFP experiments reveal  (\cite{bodin90}, \cite{jiprager}, \cite{caramana84}, \cite{japs94}) 
that $\bbE\cdot\bbB=\eta \bbJ\cdot\bbB > 0$ only at a single radius  
$0<r=r_c<a$, 
where $a$  is  the minor radius of the torus and $r_c$ is measured from the
toroidal axis. 
For $r< r_c$, $\bbE\cdot\bbB > \eta \bbJ\cdot \bbB > 0$ and for $r> r_c,\ \ $ 
$ \eta \bbJ\cdot\bbB > 0 >\bbE\cdot\bbB$.  Excluding pressure gradient and inertial terms in Ohm's law, 
such measurements imply that $\emfb_{||}\ne 0$. 
Moreover, since $\eta\bbJ\cdot\bbB-\bbE\cdot\bbB$
changes sign from negative to positive moving outward through $r_c$
(while $\bbJ\cdot\bbB$ keeps the same sign), 
$\emfb_{||}$ must 
also change sign from negative to positive across $r=r_c$.
Because the third term in (\ref{14b}) is often negligible,
(\ref{14b}) shows that the divergence of the small scale helicity
flux $\bbh$ must change sign through $r_c$.  
In the RFP, the presence of $\emfb_{||}$ is sustained by fluctuations
induced by tearing or kink mode instabilities when  the applied
$|\bbE_{||}|$ exceeds a critical value.

Because averaged quantities of this subsection remain
 functions of $r$, 
the steady dynamo just described  operates locally in $r$.
Taking the volume integral of (\ref{new8}) 
and using (\ref{new8bb})
gives
\beq
\int \emfb\cdot \bbB dV= \int{\bbh}\cdot d{\bf S}
=\int(\eta\bbJ\cdot\bbB - \bbE\cdot\bbB)dV
=\int\eta\bbJ\cdot\bbB dV + 
\int 
( {\overline \Phi} \bbB+
 {1\over2} \bbA \ts \partial_t\bbA )\cdot d{\bf S},
\label{new8b}
\ee
where we have dropped the third term of (\ref{14b})
as it is typically negligible, 
and used Gauss' theorem to convert the divergence integrals 
to a surface integrals, 
keeping in mind that for doubly connected topologies we must make sure
that $\bbh$ is analytic everywhere. This is  ensured
because our averaged quantities depend only on radius.
Accordingly, using vector identities, 
the second surface integral in (\ref{new8b}) is 
\beq
\int 
( {\overline \Phi} \bbB+
 {1\over2} \bbA \ts \partial_t\bbA) \cdot d{\bf S}
= \int_S {\overline \Phi} \bbB \cdot d{\bf S}
- {1\over2} \int_S \bbE \cdot d{\bf z}\int_S \bbA \cdot rd{\bf \hat\theta}
=\int {\overline \Phi} \bbB \cdot d{\bf S}-{1\over2}V_s\Psi_s,
\label{voltage}
\ee
where $V_s$ is the externally applied
voltage drop in the toroidal direction on the surface
of integration (applied experimentally via gaps)
and $\Phi_s$ is the toroidal magnetic flux within the surface. 
On the outer radial surface, there is no normal component of the field
and the penultimate term would vanish for that surface.
The later term of (\ref{voltage}) represents the helicity
injection. This  helicity injection is unihelical (i.e. one sign),
which is important because the magnetically driven 
dynamo relaxation can be thought of
as a process driving the injected magnetic 
helicity  the largest scale available
subject to the boundary conditions.

Because of the conducting boundaries, there is no net small scale
helicity flow through the torus.
However, it is instructive to separately 
consider the inner core  ($r<r_c$) and  shell 
($r>r_c$) regions (\cite{jiprager}). 
The discussion below (\ref{new8bb})
implies that  for  $r<r_c$ the left side of (\ref{new8b}) must be
negative. Therefore the second term of (\ref{new8b}) must also be negative, which
from the definition of $\bbh$, implies an outward flux of positive fluctuating
magnetic helicity through $r_c$.  
Analogously, positive
fluctuating helicity accumulates into the  volume defined by  $r>r_c$.
The helicity flux through $r_c$ therefore  provides the local dynamo.

Instead of dotting $\emfb_{||}$ with $\bbB$ (to get (\ref{new8})),
the dynamo is sometimes  expressed by  
dotting (\ref{14b}) with $\bbJ_{||}$. 
Using (\ref{8}) 
and  $\Lambda\equiv {\bbJ\cdot\bbB\over \bbB^2}$, this simply gives  
$\Lambda$ times (\ref{new8}), so this offers
nothing new beyond (\ref{new8b}) regarding the
local nature of the dynamo defined as a sustained $\emfb_{||}$. However,
multiplying (\ref{14b}) by $\Lambda$ and taking the volume integral,
we obtain
\beq
\int{\emfb_{||}\cdot \bbJ}dV = 
\int\Lambda{\emfb\cdot \bbB}dV = 
\int \Lambda
(\div {\bbh}-\eta\overline{\bfj\cdot\bfb} )dV = 
\int \Lambda (\eta \bbJ\cdot \bbB dV-\bbE\cdot\bbB )dV.
\label{18}
\ee
Using 
\beq
\int\Lambda\div 
{\bbh}
dV 
= \int (\Lambda \bbh )\cdot d{\bf S}-\int \bbh\cdot\nabla \Lambda dV,
\label{surf}
\ee
ignoring the third term of (\ref{18}), and eliminating $\Lambda$ in favor
of $\bbJ_{||}$, we can re-write (\ref{18}) as
\beq
\int{\emfb_{||}\cdot \bbJ}dV 
= \int (\Lambda \bbh )\cdot d{\bf S}-\int \bbh\cdot\nabla \Lambda
dV 
=
\int (\eta\bbJ_{||}^2 -\bbE\cdot\bbJ_{||}) dV,
\label{new18}
\ee
where we have used Gauss' theorem to obtain the surface integral,
just as described below equation (\ref{new8b}).
Here too the only potentially non-trivial surface terms 
are radial surface terms.
If we integrate over the full volume of the plasma (i.e. for all $r\le a$), 
then the surface term in (\ref{new18}) 
vanishes, as $\bbh$ is measured to vanish at $r=a$. 
But (\ref{new18}) then shows that a finite 
$\nabla \Lambda$ can produce a {global} dynamo effect 
 defined by  $\int_{tot}\emfb_{||}\cdot\bbJ dV\simeq
\int_{tot} \eta\bbJ_{||}^2 dV$,
where the subscript ``tot'' indicates the full volume.
The quantity $\bbh\cdot\nabla\Lambda$ gradient need not be finite everywhere 
for such an effect, only
a sufficiently non-zero $\nabla\cdot \bbh$ to ensure finite $\emfb_{||}$
and sufficiently non-zero integral in the third
term of (\ref{new18}) are required. 
For the quasi-local case,  
integration is taken over a sub-range of radii and 
either of the two middle terms of   (\ref{new18}) could dominate.
Thus $\int \bbh\cdot \nabla\Lambda dV \ne 0$ emerges as a sufficient but
not a necessary condition for a quasi-local dynamo.  This is consistent with
the actual calculation in \cite{bhattacharjee86}), 
but disagrees with  Bellan (2000) as the latter drops
the second term of (\ref{surf}).

For the global dynamo as  defined below (\ref{new18}), 
consider a case in which the magnetic helicity associated with the mean field
has a locally positive sign in the Coloumb gauge.
Then $\Lambda$ is also positive. If $\int_{tot}\eta \bbJ^2 dV$ is dominant 
on the right hand side of (\ref{new18}), then  (\ref{new18}) 
implies that the  outward flux of positive small scale helicity
(represented by $\bbh > 0$)
is anti-parallel to the direction of increasing $\Lambda$ overall.
Therefore, on average, 
 the dynamo
acts to homogenize the overall scale of  magnetic helicity.
This homogenization results 
because the magnetic helicity is injected with one sign, 
so the dynamo acts to drive the characteristic scale of magnetic twist
to the largest scale available subject to the boundary conditions.
Note that for 
the velocity driven dynamo discussed later in Sec. 3.2, kinetic helicity
rather than magnetic helicity is injected. There the dynamo 
acts instead to spectrally 
segregate magnetic helicity of opposite  signs while largely 
preserving a net zero magnetic helicity. 
In general, laboratory plasma dynamos involve the injection and evolution of 
1 sign of net magnetic helicity (unihelical) whilst kinetic
helicity  driven dynamos are bihelical.

The parallel component of the electromotive force can  be written $\emfb_{||} =\alpha\bbB$.
The pseudoscalar $\alpha$ for the laboratory case is given from (\ref{13}) and
(\ref{new8}) by
\beq
\alpha = 
{\emfb\cdot \bbB\over \bB^2}\sim 
{1\over \bB^2}\div\bbh=
{\overline{\bfe\cdot \bfb}\over \bB^2}\simeq 
{\overline{\bfe_\perp\cdot \bfb_\perp}\over \bB^2},
\label{20}
\ee 
where the last similarity follows because the fluctuations
are primarily perpendicular to the strong mean fields. 
The right side of (\ref{20}) is measured to be larger than any 
resistive contribution (\cite{japs94}),
and the values are consistent with dynamo models based on the principles
described here.
We note that the above descriptions can be extended to the case 
when global quantities are  time dependent (\cite{ji06}),
as required for more precise applications to the RFP.

\subsection{Time-Dependent, Closed, Flow-Driven Helical Dynamo}

In this subsection, instead of considering the averages to be time
averages, we consider global spatial averages
over a closed or periodic
volume such that the  surface terms vanish but the mean quantities
remain  time dependent. 
For the laboratory plasma dynamo discussed in Sec 3.1,  the electromotive 
force arises via magnetic helicity injection through a boundary term.
Here  kinetic helicity is assumed to be injected 
into the system and the boundary terms are not invoked.
These are the circumstances
considered in  recent analytic work (\cite{bf02}) and
numerical simulations (\cite{b2001}).
Another important difference between the laboratory plasma dynamo 
of the previous subsection and the flow-driven helical 
dynamo here
is that the former involves weak fluctuations on a strong mean
field, whereas the latter  involves initially weak fields
and strong fluctuations.

Following Blackman \& Field (2002), we distinguish the global volume averages from the large scale 
$k=1$ quantities, by using brackets to indicate the former and
an overbar  to indicate the latter.
For pseudoscalars, we assume the two averages
are equal, (e.g. $\overline{\bfa\cdot\bfb}=\lb\bfa\cdot\bfb\rb$).
In this case, (\ref{5a}) and (\ref{6a}) become
\beq
\partial_t\lb\bfa\cdot\bfb\rb=
-2\lb\emfb\cdot\bbB\rb
-2\eta\lb\bfj\cdot\bfb\rb
\label{5aa}
\ee
and
\beq
\partial_t\lb\bbA\cdot\bbB\rb=2\lb\emfb\cdot\bbB\rb-2\eta\lb\bbJ\cdot\bbB\rb.
\label{6aa}
\ee

In the absence of boundary terms, the magnetic helicity is gauge invariant,
but choosing 
the Coulomb gauge we can relate the current helicities to the magnetic helicities
via  $\lb\bfJ\cdot\bfB\rb =\lb k^2\bfA\cdot\bfB\rb$.
To complete the set of equations to be solved,
we need an equation for $\emfb$.
From its definition,
\beq
\partial_t\emfb=\overline{\partial_t\bfv\ts\bfb} +\overline{\bfv\ts\partial_t\bfb}.
\label{timed}
\ee
We now need equations for the fluctuating velocity and 
magnetic field $\partial_t\bfb$ and $\partial_t\bfv$. 
For $\div\bfv=0$, we have from the induction and momentum density
equations respectively, 
\begin{equation}
\begin{array}{r}\partial_{t} \bfb = \OB\cdot\nabla\bfv - \bfv\cdot\nabla\OB + 
\curl(\bfv\ts\bfb) 
-\curl\overline{\bfv\ts\bfb} +
\lambda\nabla^{2}\bfb,
\end{array}
\label{c3}
\ee
and
\begin{equation}
\begin{array}{r}
\partial_{t} {v}_q=P_{qi}({\OB}\cdot\nabla{b_i} + {\bfb}\cdot\nabla{\bB_i} 
-\bfv\cdot\nabla v_i+\overline{\bfv\cdot\nabla v_i}
+{\bf b}\cdot\nabla{b}_i-\overline{{\bf b}\cdot\nabla{b_i}})
+ \nu\nabla^{2}{v_q} 
+{f_q},
\end{array}
\label{c2}
\end{equation}
where $\bf f$ is a divergence-free 
forcing function uncorrelated with $\bfb$, $\nu$ is the viscosity, 
and $P_{qi}\equiv (\delta_{qi}-\nabla^{-2}\nabla_q\nabla_i)$
is the projection operator that arises after taking the divergence
of the incompressible momentum density equation to eliminate the 
fluctuating pressure (magnetic + thermal).
Reynolds rules (\cite{rad}) allow the interchange of brackets and time
or spatial derivatives, so the 5th term of 
(\ref{c3}) and the 4th and 6th terms in the parentheses of 
(\ref{c2}) do not contribute when put into averages and can be ignored.

The contribution to $\partial_t\emfb$ 
from the 3rd term in (\ref{timed}) can be derived 
by direct use of (\ref{c3}) in configuration space.
Following \cite{bf02}, we assume isotropy of the resulting velocity
and magnetic field correlations for terms linear in $\bbB$
and also retain the triple correlations  
The contribution 
to $\partial_t\emfb$ from the 2nd
term in (\ref{timed}) also contributes terms linear in $\bbB$, 
and triple correlations.  Here the terms linear in $\bbB$ are best
derived in Fourier space. For this, Gruzinov \& Diamond (1995) 
invoke the Fourier transform of the terms linear in $\bbB$
 contributing to $\overline{\partial_t\bfv\ts\bfb}$,  
supplemented by a linear expansion of the projection operator in $k_1<<k_2$, 
where $k_1$ is the characteristic wavenumber of the 
bracketed or mean quantities and 
$k_2$ is the characteristic wavenumber of the fluctuating quantities
$\bfb$ and $\bfv$. Collecting all surviving 
terms, we then have for (\ref{timed})
\beq
\begin{array}{r}
\partial_t{\emfb}=
{1\over 3}(\overline{\bfb\cdot\curl\bfb}
-\overline{\bfv\cdot\curl\bfv})\bbB 
- {1\over 3}\overline{
\bfv^2}\curl\bbB
+\nu\overline{\nabla^2\bfv\ts\bfb} +\lambda\overline{\bfv\ts\nabla^2\bfb}
+ {\bf T}^V+{\bf T}^M,
\end{array}
\label{timed33}
\ee
where ${{\bf T}^M}=\overline{ \bfv\ts\curl(\bfv\ts\bfb)}$ 
and ${T_{j}^V}=\overline{(\ep_{jqn}\lb P_{qi}(\bfb\cdot\nabla b_i
-\bfv\cdot\nabla v_i) b_n}$
are  the triple correlations. Note that 
the 3rd, 4th, 6th and 8th  
terms in (\ref{timed33})
 come from  
the $\overline{\bfv\ts\partial_t\bfb}$ term of (\ref{timed})
and the 2nd 5th and 7th terms come from the 
$\overline{\partial_t\bfv\ts\bfb}$ term of (\ref{timed}).
For  $\emfb_{||}$ 
we have
\beq
\partial_t\emf_{||}=(\overline{
\partial_t\bfv\ts\bfb}
 +\overline{\bfv\ts\partial_t\bfb})\cdot\bbB/|\bbB|+
\overline{\bfv\ts\bfb}\cdot\partial_t(\bbB/|\bbB|).
\label{timedp}
\ee
Substituting (\ref{timed33}) into 
(\ref{timedp}) gives 
\beq
\partial_t\emf_{||}= {\tilde\alpha}{\bbB^2/|\bbB|}
-{\tilde\beta}{\bbB\cdot\curl\bbB}/|\bbB|-{\tilde \zeta}\emf_{||}
\label{2emf}
\ee
where 
${\tilde\alpha}
=(1/3)(\overline{\bfb\cdot\curl\bfb}-\overline{\bfv\cdot\curl\bfv})$,  ${\tilde\beta} = (1/3)\overline{ \bfv^2}$, and  $\tilde \zeta$ 
accounts for microphysical dissipation
terms, the last term of (\ref{timedp}), and
${\bf T}^M + {\bf T}^V\ne 0$. 

The incorporation of the triple correlations via $\tilde \zeta$ 
was subsequently named the ``minimal $\tau$'' 
closure  (\cite{bf03}). In reality, $\tilde \zeta$ can
be a function of wavenumber when spectral models are considered
(\cite{kr96}). However, the 
simple minimal $\tau$ closure is  an
improvement over the  
first order smoothing approximation (\cite{moffatt,parker,vishniac})
in which triple correlations are ignored, and simpler than the
eddy damped quasi-normal Markovian closure (\cite{pfl}). 
In general, triple correlations
cannot be ignored because flow-driven dynamos 
involve fluctuations which are generally not 
 small compared to the mean magnetic field. Moreover,  it is these
very triple correlation terms which lead to a turbulent cascade, and
the presence of a turbulent spectrum. The efficacy of the minimal $\tau$ 
closure used for passive scalar diffusion by \cite{bf03}
was verified numerically (\cite{branfluid}), and shows
reasonable agreement with numerical tests in the MHD regime (\cite{bs05b}).

The equations to be solved for the closed, helical
flow-driven dynamo case here
 are
(\ref{5aa}), (\ref{6aa}), and 
(\ref{timedp}). The large scale field which grows is fully helical,
and so the magnetic energy and magnetic helicity equations are essentially
the same. The resulting dynamo is a modern representation of the $\alpha^2$
dynamo (e.g. \cite{moffatt}) that  incorporates the dynamical backreaction of the
magnetic field on the kinetic helicity driving the flow
and the conservation of magnetic helicity. 
Solutions of these three equations agree with simulations 
as described in (\cite{bf02}). 
The essence of the dynamo growth is as follows: the initial
system is assumed to be driven with a finite 
$\overline{\bfv\cdot\curl\bfv}\simeq
\lb\bfv\cdot\curl\bfv\rb$.
This grows a finite $\emfb_{||}$ which then grows large scale 
magnetic helicity via (\ref{5aa}). But due to magnetic helicity
conservation, as seen in (\ref{6aa}), 
the small scale magnetic helicity $\overline{\bfa\cdot\bfb}$ 
must then grow in opposite sign to that of the large scale. 
This also grows the small scale current helicity $\overline{\bfj\cdot\bfb}$
of the same sign. This in turn quenches $\tilde\alpha$. The system
reaches a steady state: If the decay of the large scale helicity 
were to supersede growth, the small scale helicity would also 
deplete, and growth of the large scale field would again begin
if $\overline{\bfv\cdot\curl\bfv}$ is steadily driven.

In the traditional  $\alpha^2$ dynamo of the standard texts (\cite{moffatt}),
the equation for the large scale field is solved, with an
imposed form of the electromotive force and the linear growth
equation results. \cite{bb03} emphasize that 
magnetic helicity is neither conserved in the equations nor the diagrams
of the dynamo in the standard texts.
 In the modern version just discussed, 
the large scale field evolution equation
was replaced by the large scale magnetic helicity evolution
equation and the additional time dependent equations for
small scale helicity evolution and turbulent electromotive
force evolution are coupled into the theory dynamically.

It should be noted that the two-scale analytic 
approach has been generalized to a 4 scale approach (Blackman 2003)
to assess whether the small scale 
magnetic helicity tends toward the dissipation scale or toward
the forcing scale (with the large scale magnetic helicity 
migrating toward the even larger box scale.)
The analysis shows that the small scale magnetic helicity first
appears toward the resistive scale but migrates toward the forcing
scale before the end of the kinematic regime.
Numerical simulations of helical dynamos
in a periodic box (Brandenburg 2001; Maron \& Blackman 2002)
also show that the magnetic helicity in saturation
peaks with opposite signs  at the forcing scale and
box scale respectively. This is a bihelical dynamical equilibrium state in which
the small and large scale magnetic helicities both
migrate to the largest scales available to them. The driving
kinetic helicity ensures that these two scales are distinct
and prevents the small scale magnetic helicity from migrating 
to the box scale.

The dynamical backreaction 
approach of this section can be generalized to an $\alpha-\Omega$ dynamo.
Then the large scale magnetic helicity evolution
equation must be replaced by the vector equation for the large
scale field itself, but the small scale and turbulent electromotive
force equations are also included dynamically (\cite{bb02}).

\subsection{Steady State, Open Flow-Driven Helical Dynamo and Implications for
Coronal Magnetic Relaxation }

Consider now the limit of (\ref{5a}) and 
(\ref{6a}) in which the time evolution and resistive terms
are ignored, but the divergence
terms are kept. We then have respectively, 
\beq
0=2\lb\emfb\cdot\bbB\rb
-\div\lb{\overline\Phi} \bbB + \bbE\ts \bbA\rb
\label{26}
\ee
and 
\beq
0=
-2\lb\emfb\cdot\bbB\rb
-\div\lb{\phi} \bfb + \bfe\ts \bfa\rb
\label{27}
\ee
Combining these two  equations reveals that  the fluxes
of large and small scale helicity through the system
boundary are equal and opposite. This has important implications
for a helical flow-driven dynamo inside  an
astrophysical rotator: If  
 helical motions were to sustain  kinetic helicity
inside of the rotator,  large and small scale
magnetic helicities of opposite sign grow as discussed in the previous subsection.
The existence of a steady-state with open boundaries
implies that the  boundary fluxes of magnetic helicity contribute to 
the respective loss terms 
in the large and small scale magnetic helicities. The
corona would be supplied with bihelical 
structures (\cite{bf00b},\cite{bb03},\cite{bb03b}). This is 
consistent with time-averaged steady coronae of the sun (e.g. \cite{sz}) and AGN accretion disks 
(e.g.Haardt \& Maraschi 1993; Field \& Rogers 1993).
The bihelical nature of the field supplied by the dynamo
and the sign dependence of the injected helicity on whether
surface shear operates on a scale larger or smaller than that of
a given loop's footpoint separation are reasons why
extracting the dominant sign of the solar coronal magnetic
helicity in each hemisphere of the sun has been somewhat elusive
(e.g. Demoulin et al 2002).

The evolution of magnetic structures 
 injected into a corona is conceptually analogous to the evolution
 of a 
magnetically dominated laboratory plasmas to injection of magnetic helicity
such as in   a Spheromak (e.g. \cite{bellan}, Hsu \& bellan 2002).
Even though the corona in the astrophysical case  receives
 injected helicity of both signs, 
the guiding principles understood from laboratory plasmas
are  applicable.  The  experiment of Hsu \& Bellan 2002 
provides a direct analogy to helical loops of flux rising
into an astrophysical corona from its rotator below.
The loops coalesce at the symmetry axis and form
a magnetic tower. For large enough helicity injection, 
the tower can break off a Spheromak blob from the kink instability. 
An astrophysical corona can also be modeled as a statistical aggregate 
of magnetically loops and 
the corona can be thought of as a single dynamical entity (\cite{bf04}).
The helicity flux to the corona in astrophysical rotators
acts  as a seed for subsequent magnetic relaxation therein.
The relaxation opens 
field lines that the form coronal holes or jets.

\subsection{Time-dependent, Open, Flow-Driven Helical Dynamos}

In general, both the time dependent and the  flux terms 
in equation (\ref{5a}) and (\ref{6a}) should be included dynamically.
We briefly describe two calculations of flow-driven dynamos
which incorporate both, using different sets of approximations.
In this section we assume the that overbars indicate spatial averages.

In the context of the Galaxy, Shukurov et al. (2006)  have 
solved the mean field induction equation for $\bbB$ with
$\emfb_{||}$ determined from setting 
$\partial_t \emfb_{||}=0$ in (\ref{2}).
The $\emfb_{||}$ involves the difference between the kinetic and current 
helicities which can be related to
small scale magnetic helicity in the Coulomb gauge.
(Shukurov et al. (2006) formally use a gauge invariant
helicity density, derived by Subramanian \& Brandenburg (2006)
to replace the use of the magnetic helicity density but
the key role of the boundary terms is conceptually independent of this.)  
Effectively, 
Shukurov et al. (2006) therefore solve the induction  equation for $\bbB$, 
(which depends on $\emfb_{||}$ and thus $\lb\bfa\cdot\bfb\rb$)
and Eq. (\ref{6a}) for $\lb\bfa\cdot\bfb\rb$.
The divergence term in (\ref{6a}) can be replaced with one of the form 
$\propto \nabla\cdot (\lb\bfa\cdot\bfb \rb \bbV)$, where $\bbV = (0,0,\bV_z) $ is the mean
velocity advecting the small scale helicity out of the volume.
This mean velocity also appears in the induction equation for $\bbB$, 
highlighting that the  loss terms in the small scale helicity equation also
imply advective loss of mean field.
This approach supports the concept
(\cite{bf00a}) 
that a flow of small scale helicity toward the boundary may help
to alleviate the backreaction 
of the small  scale  magnetic helicity on the 
kinetic helicity which drives the dynamo in  $\emfb_{||}$. 
However, if $\bV_z$ is too large, it may
carry away too much of the  mean field which the dynamo is trying to grow
in the first place.
In general, more work is needed to calculate $\bV_z$ from
first principles, and its effect on large and small scale fields.
Coupling the dynamo in the rotator to the magnetic helicity
evolution in the corona above is also of interest.

A more restrictive time dependent dynamo that includes boundary terms, maintains the time dependence in (\ref{5a}), but implicitly
assumes that equation (\ref{6a}) reaches a steady-state has also been
studied (\cite{vishniac}).  This approach  explicitly
incorporates the role of shear into the helicity flux.
Although the approach involves assumptions that have 
now been avoided in more general calculations of helicity fluxes 
(e.g. Subramanian \& Brandenburg 2004) (one being the the first order smoothing approximation which can be avoided by the ``minimal tau'' closure 
discussed in section 3.2), 
the Vishniac \& Cho (2001) paper 
identifies how a time dependent flow-driven dynamo 
in a Keplerian shear flow might be sustained by a  magnetic helicity flux.
We choose to outline this approach in more detail here 
in part because the Vishniac \& Cho (2001) is  less 
transparent than the more recent paper of (\cite{shukurov06}) but 
did first explicitly use the boundary flux to solve a time dependent 
dynamo equation. In consolidating  the calculation here,
we show how it dovetails into the unified framework of section 2.

If we ignore the resistive terms, the relevant forms of 
(\ref{5a}) and (\ref{6a})  are
\beq
\partial_t (\bbA\cdot\bbB)= 
2\emfb\cdot\bbB
-\div({\overline\Phi} \bbB + \bbE\ts \bbA)\
\label{26a}
\ee
and 
\beq
0=
-2\emfb\cdot\bbB
-\div(\overline{{\phi} \bfb} + \overline{\bfe\ts \bfa})
\label{27a}
\ee
Using (\ref{27a}),  $\emfb_{||}$ 
can be directly written in terms of the  
small scale helicity flux as 
\beq
\emfb_{||}=-{\bbB\over B^2}
\nabla\cdot(\overline{{\phi} \bfb} + \overline{\bfe\ts \bfa})
=
-{\bbB\over B^2}
\nabla\cdot \overline{(-\nabla\phi +\bfe)\ts\bfa}.
\label{vc1}
\ee
\cite{vishniac} then use
(\ref{vc1})  
in the equation for the mean magnetic field applied to an accretion disk
whose mean quantities are axisymmetric. The mean field equation is 
\beq
\partial_t\bbB=\curl \emfb +\curl (\bbV\ts\bbB)+\lambda \nabla^2\bbB.
\label{vc2}
\ee
Solving (\ref{vc2}) requires use of (\ref{vc1}).  
\cite{vishniac} invoke a correlation time $\tau_c$ such that  $\bfa \simeq -(\bfe +\nabla\phi)\tau_c$, 
(note: they define $\bfe_{mf}\equiv -\bfe$ and work with $\bfe_{mf}$).  
This reduces the last term of (\ref{vc1}) to 
\beq
2{\bbB\over B^2}
\nabla\cdot \overline{\bfe\ts\nabla\phi}\equiv -{\bbB\over B^2}\nabla \cdot {\bf J}_H,
\label{helcur}
\ee
which defines the helicity flux $\bfJ_H$. (\cite{vishniac} are 
missing a factor of 2).

To proceed, $\phi$ can be inverted in terms of $\bfe$ 
by Fourier transform, which gives
\beq
J_{H,i}
=2\int {d^3{\bf r}\over 4\pi r}\ep_{ijk}
\overline{
e_j({\bf x})\partial_k\partial_le_l({\bf x}+{\bf r})},
\label{vc16}
\ee
where the spatial derivatives operate on $\bf r$ and 
an assumption of isotropy in wavenumber (not explicitly stated in 
\cite{vishniac} was used to obtain this).
The notoriously troubling but useful
first order smoothing approximation, where 
terms nonlinear in fluctuating quantities are ignored (see section
3.2 above) was then used for $\bfe$
to obtain $\bfe \simeq -\bfv \ts\bbB-\bbV\ts \bfb$. 
Ignoring $\bbV$ here in the equation for $\bfe$ (not necessarily justified) 
and  assuming ${\bf r}<< {\bf x}$
Eq. (\ref{vc16}) becomes, after some algebra  
\beq
J_{H,i}\sim
2l^2 \tau_c \overline{ e_j({\bf x})\partial_k\partial_l e_l({\bf x})}
= 2l^2 \tau_c\ep_{lts}(\bB_n\bB_t 
\overline{v_i({\bf x})\partial_n\partial_lv_s({\bf x})}
- \bB_i\bB_t 
\overline{ v_k({\bf x})\partial_k\partial_lv_s({\bf x})}),
\label{33}
\ee
where $l$ is a spatial correlation scale.
Assuming incompressible flow for the fluctuations, 
and assuming that total spatial derivatives
of the velocity correlations are small, the third term on the
right of (\ref{33}) vanishes and integrating the last term by
parts gives
\beq
J_{H,i}\sim 2l^2 \tau_c \overline{\bbB\cdot\nabla v_i\bbB\cdot {\curl \bfv}}.
\label{34}
\ee
This current  is then used in (\ref{vc2}) to allow growth
of $\bbB$. 
Subramanian \& Brandenburg (2004) and  
Brandenburg \& Subramanian (2005) show that this is one of a number
of current terms that emerge in a more general calculation which avoids
the first order smoothing approximation (see section 3.2) 
with additional fluxes arising when $\bbV$ is included in $\bfe$.
Nevertheless, the importance of  Eq. (\ref{34}) 
is that it, in principle, allows  dynamo growth of  the large scale field to be driven entirely by the small scale helicity flux without any
kinetic helicity. 
Determining whether this
works in practice needs further work.

The role of the Vishniac-Cho flux  
has been investigated numerically in a few experiments 
with somewhat mixed results:
Brandenburg \& Subramanian (2005a) 
found that the flux can sustain the field growth in the
absence of kinetic helicity but only if the field already
exceeds  $\sim 70\%$ of the equipartition value of the turbulent field. 
On the other hand, when kinetic helicity is present, the Vishniac-Cho flux has
been numerically shown (\cite{b2005})  to alleviate catastrophic
quenching from small scale helicity build up
by allowing ejection of small scale helicity
through the boundary, consistent with conceptual 
suggestions of the role of the boundary terms (e.g. Blackman \& Field 2000a). 

The role of the boundary flux may be particularly important
when a rapid, unquenched, cycle period is involved such as in the Sun (Blackman \& Brandenburg 2003). 
However, as also emphasized in section 3.3, 
large scale helicity flux likely accompanies any small scale helicity
flux. Significant loss of the large scale field would imply
removal of the large scale field that the dynamo is invoked
to generate inside the rotator thereby lowering its maximum value
inside the rotator. Care in identifying the relative amount of 
large and small scale helicity flux is warranted.
The same issue also arises in the Galactic field calculation of Shukurov et al. (2006).

A flux driven helical dynamo 
is conceptually  distinct from the closed volume 
kinetic helicity driven dynamo discussed in section 3.2. 
The flux driven dynamo is more analogous to the laboratory plasma dynamo
discussed in section 3.1 in the steady state approximation, but with a key difference: The source of the helicity flux for the velocity driven
case  depend on fluctuations driven by, and coupled to, the velocity shear 
(e.g. Balbus \& Hawley 1998), not fluctuations
driven by current driven magnetically dominated  instabilities.

\section{Insights from the Unified Framework}

To exemplify the benefit of the unified approach to helical dynamos,
we consider insights that emerge from synthesizing the cases
discussed in the previous sections. In particular,
we consider what can be learned about 
saturation, dissipation vs. boundary terms, and 
dynamical equilibria vs relaxed states.
As mentioned in the Sec. 1, nonlinear saturation and the
extent to which helical dynamos depend on dissipation has been
a long standing topic of study, particularly for flow driven helical dynamos.
The magnetically driven example of section 3.1  
can actually provide insight  on the flow driven cases.

First, the fact that the large scale magnetic field
dominates the fluctuating field in section 3.1 highlights that there
is no universal requirement that the mean field energy in the steady state
be less than that of the fluctuating magnetic field for a working helical 
dynamo. This highlights that the Zeldovich relations (Zeldovich 1956;
Zeldovich  et al. 1983) relating the mean and fluctuating fields
when a uniform field is imposed and the small scale field is amplified non-helically, were never derived as a constraint on helical dynamos (Blackman \& Field 2005).
Second, the laboratory plasma experiments show that the 
resistive terms are small in the RFP dynamos.  Eq. (\ref{20}) reveals that the
consistent explanation for the sustenance of the mean magnetic field
via $\emfb_{||}$
in an  RFP dynamo  is the flux of small scale magnetic helicity. 
In a steady-state, the field is neither growing nor decaying but
observed RFP sawtooth oscillations occur on a faster-than-resistive
time scale and can be accommodated by (\ref{20}) when augmented by the term 
${1\over \bB^2}\partial_t{\overline{\bfA\cdot\bfB}}$  from 
the left side of (\ref{5a}).  As discussed in the previous section,
flux terms can allow the dynamo to incur cycles or oscillations 
 on time scales not limited by resistivity.
Note the distinction between  section 3.1 and section 3.2. In the latter, 
we consider a closed volume time-dependent, flow driven 
helical dynamo. As described therein, the 
growth starts out fast and formally independent of the magnetic dissipation
rate. Then as the small scale scale magnetic helicity builds up,
the dynamo slows to become resistively limited and a steady-state balance
occurs between the electromotive force and the resistive terms. But,
 unlike the case of section 3.1, the helicity flux terms vanish in section 3.2. 
Therefore the only terms that can balance the electromotive force in the steady
state are the resistive terms for section 3.2.

The role of the helicity flux  for a flow driven dynamo is further exemplified
in the cases of section 3.3 and 3.4.
When the averages
are taken locally, the helicity flux terms  for a flow driven
dynamo can dominate resistive terms.
In an open astrophysical object like a star or accretion disk 
the flux terms  send magnetic helicity and magnetic energy to the corona.
Indeed for  either the time dependent case of section 3.4 or the steady state
of 3.3, the electromotive force is balanced 
by a combination of time evolution of magnetic helicity and boundary
terms. In both of these cases the dissipation terms are negligible.

Finally, a further comment on the relation 
between dynamical equilibria and fully relaxed states.
When kinetic helicity is steadily injected 
for a velocity driven dynamo or when  magnetic helicity
is steadily injected for a magnetically driven dynamo,
an equilibrium or a quasi-steady state can be reached. 
The latter is 
exemplified by sawtooth oscillations in an RFP dynamo or as the 22 year solar 
dynamo cycle . The driving or injection
take the system away from the relaxed state, but the dynamo,
fueled by induced instabilities, 
continually competes to  take the system back toward the relaxed state.
An important difference between injecting one sign of magnetic helicity 
vs. one sign of kinetic helicity is that the latter leads to a 
spectral or spatial segregation of helicities of opposite sign.
(the bihelical dynamical equilibrium for the $\alpha^2$ dynamo discussed in 
section 3.2). When magnetic helicity of one sign is injected,
that magnetic helicity seeks the largest scale available to it.
When the driving is turned off and the resistivity small, 
the system can  fully relax, but eventually the field decays.  
The case of a forced vs. decaying magnetically dominated turbulent 
dynamo was studied in Blackman \& Field (2004).
Numerical simulations of the relaxation of helical MHD turbulence
with initially equal kinetic and magnetic 
energy densities were studied in (\cite{cb01}). These
works  show the tendency
of a single sign of magnetic helicity to migrate toward the largest
scale for magnetic helicity injected with a single sign.

\section{Summary and Conclusion}

We have used the equations for magnetic helicity evolution as a 
unifying  framework  for helical dynamos and we have discussed 
both magnetically driven laboratory plasma dynamos and flow-driven astrophysical 
dynamos within this framework. We summarize the common principles
and distinguishing features of these dynamos here.

Laboratory plasma helical dynamos typically involve
a magnetically dominated initial state with a dominant mean magnetic toroidal 
magnetic field. When an external toroidal electric field is applied,
a current is driven along the field which injects magnetic helicity
of one sign into the system and generates a poloidal  field.
For sufficiently strong applied electric fields, the system is driven
sufficiently far away from its relaxed state that 
helical tearing or kink mode instabilities occur.  The resulting 
fluctuations produce a turbulent electromotive 
force that drives the system back toward the relaxed state.  
The relaxed state for a such a unihelical dynamo 
is the state in which the magnetic helicity
is at the largest scale possible, subject to boundary conditions.
When the unihelical helicity injection is externally sustained for a real 
system, a dynamical
equilibrium with oscillations can incur as  the system evolves toward and away
from the relaxed state. The time-averaged  
electromotive force is maintained by a spatial (radial) flux of small scale 
magnetic helicity within the plasma. 
The injection of helicity is balanced by the dynamo relaxation
so the  dynamo sustains the field configuration  against decay.

Like the laboratory plasma dynamos, the flow-driven helical dynamos
require a turbulent electromotive force to grow or sustain magnetic helicity
at large scales. These dynamos are often invoked as an explanation for the 
large scale fields of astrophysical rotators.
Unlike the laboratory plasma dynamos, for the canonical flow-driven helical 
dynamo, the initial mean field is weak and the velocity fluctuations  are strong.  For the simplest time-dependent case in a closed volume, the 
electromotive force is proportional to the difference between
the current helicity and the kinetic helicity densities.
The latter initially dominates and this drives growth of the large
scale magnetic helicity by sending one sign of the magnetic helicity
to large scales and the other sign 
to scales at or below the scale of the dominant velocity fluctuations.
Here, unlike the laboratory plasma dynamo, no magnetic helicity is injected
and so the dynamo acts to segregate magnetic helicity of opposite signs
spatially or spectrally.
The build up of the small scale magnetic helicity also grows the small
scale current helicity which offsets the kinetic helicity contributions
to the electromotive force and quenches the  dynamo into a steady state.
In the absence of boundary terms, the steady state is one in which
growth is balanced by resistive dissipation. 
When boundary flux terms are allowed, the electromotive force can be
sustained by a magnetic helicity flux, just as in the laboratory case.
Such a helicity flux can arise from a Keplerian velocity 
shear and stratification rather than the current driven instabilities of 
laboratory plasma dynamos.

In the astrophysical case, when the resistive terms in the magnetic helicity evolution equation 
are  small,   equal in magnitude (but opposite in sign) small  and 
large scale fluxes of magnetic helicity are injected to coronae in the steady
state.  For a time dependent situation, open 
 boundaries can, in principle, allow the dynamo to 
overcome any long term resistive saturation by ejecting the 
offending small scale helicity from a volume of interest. However,
such ejection will undoubtedly involve ejection of large scale
field as well, which can  reduce the steady state saturation value
of the large scale field inside the rotator volume 
compared to the asymptotic saturation of the closed case.
 
In the magnetically dominated 
corona, energy associated with this helicity can be
extracted into  high energy particles.
Astrophysical coronae, with
their underlying rotators acting as magnetic helicity injectors,
are the astrophysical circumstance most directly
analogous to the magnetically driven dynamo physics of laboratory plasma
devices such as Spheromaks and RFPs. In particular, a single loop 
with footpoints being sheared or twisted provides a direct analogy to
the precursor to Spheromak formation. Two subtle differences that arise
when considering  the analogy 
between coronae and laboratory plasmas more carefully 
are that (1) the corona is composed
of many injection sites each analogous to a Spheromak,
and that (2) the dynamo in the rotator below injects both signs of magnetic
helicity into the corona.
Nevertheless, relaxation of magnetic fields in  astrophysical coronae can produce the very largest scale fields associated with coronal holes and jets
and the principles of laboratory plasma dynamos are applicable.

Our goal  has been to provide a base that  anchors common
principles and differences between laboratory and 
astrophysical helical dynamos to foster further cross-disciplinary work.
We have avoided a detailed  
exposition about each type of dynamo
in specific laboratory and astrophysical systems
in order to  focus on the basic concepts.

{\bf Acknowledgments}:
We would like to thank A. Brandenburg, 
G. Field, S. Prager, and E. Vishniac for 
helpful discussions and comments.
We acknowledge the stimulating meetings organized by
Center of Magnetic Self-organization in Laboratory and Astrophysical Plasmas
as well as the Isaac Newton Institute for Mathematical Sciences, University of
Cambridge, under EPSRC Grant N09176.
EGB acknowledges support from 
NSF grants AST-0406799, AST-0406823 and NASA grant ATP04-0000-0016.
HJ acknowledges support of DoE through contract  DE-AC02-76-CH03073.

\end{document}